\documentclass[12pt]{article}
\usepackage{setspace}

\begin{document}

\title{The $2(2S+1)$-- Formalism and Its Connection with Other Descriptions}

\author{Valeriy V. Dvoeglazov\\
Universidad de Zacatecas, 
Ap. Postal 636, Suc. 3\\
Zacatecas 98061, Zac., M\'exico\\
E-mail: valeri@fisica.uaz.edu.mx}

\date{\empty}

\maketitle

\begin{abstract}
In the framework of the Joos-Weinberg $2(2S+1)$-- theory for massless
particles,  the dynamical invariants have been derived from the Lagrangian
density which is considered to be a 4-- vector. A la Majorana
interpretation of the 6-- component ``spinors`", the field operators of
$S=1$ particles, as the left-- and right--circularly polarized radiation,
leads us to the conserved quantities which are analogous to those obtained
by Lipkin and Sudbery. 
The scalar Lagrangian of the Joos-Weinberg theory is shown to be equivalent
to the Lagrangian of a free massless field, introduced by Hayashi. As a
consequence of a new ``gauge" invariance this skew-symmetric field describes
physical particles with the longitudinal components only.
The interaction of the spinor field with the Weinberg's $2(2S+1)$- component
massless field is considered. New interpretation of the Weinberg field function is proposed. 
\end{abstract}

In the beginning of the sixties Joos~\cite{Joos}, Weinberg~\cite{Weinberg}
and Weaver, Hammer, and Good~\cite{Weaver} developed 
free particle theories for arbitrary spins $S=0,{1 \over 2}, 1,{3 \over 2}
\cdots$ on using the Wigner's ideas~\cite{Wigner} of construction of the
quantum field theory. Following this description, the spin-one case
{}~\cite{Sankaranarayanan}-\cite{Tucker} as well as the spin-${3 \over 2}$
case~\cite{Shay} have been presented. The formulas for the
Hamiltonian for any spin have also been obtained~\cite{Mathews,
Williams}\footnote{ I would also like to mention the following earlier articles
concerning with this formalism~\cite{Santos}-\cite{Dvoeglaz}.}
The field functions in this approach form the basis
 of the $(S,0)\oplus(0,S)$ representation of the Lorentz group. They are
presented by the $2(2S+1)$-- component ``spinor":
\begin{equation}\label{eq:psi}
\Psi=\left (\matrix{
\chi_\sigma\cr
\phi_\sigma\cr
}\right ),
\end{equation}
The transformation rules
\begin{eqnarray}
\cases{\chi_{\sigma}(\vec p)=exp\left (+\theta \hat{\vec p}\hat{\vec J}\right )
\chi_{\sigma}\left (0 \right ), & $ $\cr
\phi_{\sigma}\left (\vec p\right )=exp\left (-\theta\hat{\vec p} \hat {\vec
J}\right ) \phi_{\sigma}\left (0 \right ) & $ $}
\end{eqnarray}
(with $\theta$ is the boost parameter, $tanh \theta=\frac{\mid\vec p\mid}
{E}$, $\hat{\vec p}=\frac{\vec p}{\mid \vec p \mid}$, $\hat{\vec J}$ is the
angular momentum operator) represent the generalizations of the
well--known Lorentz boosts for the Dirac particle. 
It was noted in
Ref. [2b, p. 888] that the equation for this ``spinor":
\begin{equation}
(\gamma_{\mu\nu}p_\mu p_\nu + m^2)\Psi=0
\end{equation}
can be transformed to the equations for left-- and right--circularly
polarized radiation when the massless $S=1$ field being considered.
The $\gamma_{\mu\nu}$  matrices are  covariantly defined $6\otimes6$-
matrices~\cite{BM}, $\mu,\nu=1\ldots 4$.

Thus, we come to the Maxwell's free-space equations (Eqs. (4.21) and
(4.22) of Ref. [2b]):
\begin{eqnarray}\label{eq:Maxwell}
\cases{\vec\nabla\times[\vec E-i\vec H]+i(\partial/\partial t)[\vec E-i\vec
H]=0,& $ $\cr
\vec\nabla\times[\vec E+i\vec H]-i(\partial/\partial t)[\vec E+i\vec H]=0,& $
$}
\end{eqnarray}
in vacuum provided that we consider (\ref{eq:psi}) as the ``bivector"\,\footnote{See also
{}~\cite[p.149]{Ohnuki} for discussion about interpretations of components
of the field transforming on the $(S,0)\oplus (0,S)$ representation of the
Lorentz group.} which can be decomposed as, e.g.~\cite{Recami}:
\begin{eqnarray}\label{eq:EH}
\cases{
\chi=\quad\vec E+i\vec H,& $ $\cr
\phi=-\vec E+i\vec H& $ $}
\end{eqnarray}
($\vec E$ and $\vec H$ are the 3-vectors). In fact, this is the formulation which is
similar to~\cite{Recami2}-\cite{Gianetto}\footnote{See~\cite{Collins} for
discussion about connection of
$2(2S+1)$-- component multispinor $\Psi=(\Psi_{\alpha_1\ldots\alpha_{2S}})$,
of the massless Bargmann-Wigner equations with the antisymmetric
field tensor $F_{\mu_1\nu_1\ldots\mu_S\nu_S}$.}.

Attempts at describing the quantized electromagnetic field in the terms of
electric and magnetic field vectors $\vec E, \vec H$ (but not potential) as
independent
variables, or, equivalently, antisymmetric strength tensors, have been
undertaken
previously~\cite{Anderson}-\cite{Sudbery}. For example, in Ref.~\cite{Sudbery}
the 4--vector Lagrangian density:
\begin{equation}
{\cal
L}_{\alpha}=^{*}F^{\mu\nu}\partial_{\nu}F_{\mu\alpha}-F^{\mu\nu}\partial_{\nu}\,^{*} 
F_{\mu\alpha}-2^{\,\, *} F_{\alpha\mu}j^{\mu}
\end{equation}
($F_{\mu\nu}$ is the electromagnetic field tensor, $^* F_{\mu\nu}=\epsilon
_{\mu\nu\rho\sigma}F^{\rho\sigma}$ is its
dual, $j^{\mu}$ is the electromagnetic current 4--vector) has been used to
determine the new conserved quantities analogous to those  deduced
from the Lipkin tensor~\cite{Lipkin}. The remarkable feature of this
formulation is that the  energy-momentum conservation is associated {\it not}
with the translational invariance but with the invariance under duality
rotations.

In the present article the similar properties are shown for the Lagrangian
density of the Joos-Weinberg theory. Following ~\cite{Sudbery},
the Lagrangian is chosen to be the 4--vector\footnote{See~\cite{Fushchich}
for the details of the vector Lagrangian description.}:
\begin{equation}\label{eq:Lagr}
{\cal L}_{\alpha}=-i\bar\Psi\gamma_{\alpha\beta}\partial_{\beta}\Psi
+i(\partial_{\beta}\bar\Psi)\gamma_{\alpha\beta}\Psi.
\end{equation}

On using the variational principle of the stationary action the above Lagrangian
leads to the Euler--Lagrange equations:
\begin{eqnarray}\label{eq:LagEul}
\cases{\gamma_{\alpha\beta}\partial_{\beta}\Psi=0,& $ $\cr
(\partial_{\beta}\bar\Psi)\gamma_{\alpha\beta}=0,& $ $}
\end{eqnarray}
which are, in fact, the Eqs.($4, 4'$) of Ref.~\cite{Recami2}. When $\alpha=4$
Eqs.(\ref{eq:LagEul}) are rewritten to  Eqs. (\ref{eq:Maxwell}), whereas when
$\alpha=i=1,2,3$ we come to:
\begin{eqnarray}
\cases{
\epsilon_{ikl}\frac{\partial E_l}{\partial t}+\partial_k H_i-
\partial_i H_k+(\partial_j H_j)\delta_{ik}=0,& $ $\cr
\epsilon_{ikl}\frac{\partial H_l}{\partial t}+\partial_i E_k-
\partial_k E_i - (\partial_j E_j)\delta_{ik}=0,& $ $}\,.
\end{eqnarray}
The symmetric and antisymmetric parts give us the usual four Maxwell's
equations. Let us mark the coincidence of these equations with Eqs.
on p.L34 of  Ref.~\cite{Sudbery} as well as with the system of equations (17)
in Ref.~\cite[p.76]{Jancewicz}:
\begin{eqnarray}
\cases{
\frac{\partial \hat{H}}{\partial t}+\vec\nabla\wedge\vec E - (\vec\nabla\vec
E)\delta_{ik}=0,& $ $\cr
\frac{\partial\hat{E}}{\partial t}-\vec\nabla\wedge\vec H+
(\vec\nabla\vec H)\delta_{ik}=0.& $ $}
\end{eqnarray}
Here hats above $E$ and $H$ designate volutors.

The use of the proposed Lagrangian (\ref{eq:Lagr}) simplifies the calculations. 
It gives us the opportunity to
obtain dynamical invariants:

1)The energy-momentum tensor has the following form:
\begin{equation}
T^{\mu\nu}_{\alpha}={\cal L}_{\alpha}\delta_{\mu\nu}+i\bar\Psi\gamma_
{\alpha\nu}\partial_{\mu}\Psi-i(\partial_{\mu}\bar\Psi)\gamma_{\alpha\nu}
\Psi.
\end{equation}

2)The angular momentum tensor is
\begin{eqnarray}
\lefteqn{{\cal M}^{\nu ,\mu\beta}_{\alpha}=x_{\mu}T^{\nu\beta}_{\alpha}-
x_{\beta}T^{\nu\mu}_{\alpha}+}\nonumber\\
&+&i\bar\Psi\gamma_{\alpha\nu}A^{\Psi}_{\mu\beta}\Psi-i\bar\Psi A^{\bar\Psi}
_{\mu\beta}\gamma_{\alpha\nu}\Psi
\end{eqnarray}
(with $A^{\Psi}_{\mu\beta}$ and $A^{\bar\Psi}_{\mu\beta}$ are the generators
of the Lorentz transformations).\\
And, finally,

3)the current tensor is equal to
\begin{equation}
J^{\mu}_{\alpha}=-2\bar\Psi\gamma_{\alpha\mu}\Psi\,.
\end{equation}
It is obtained as the consequence of gradient transformations:
\begin{eqnarray}
\cases{
\Psi=e^{i\theta}\Psi,& $ $\cr
\bar\Psi=\bar\Psi e^{-i\theta}& $ $}\,,
\end{eqnarray}
where $\bar\Psi=\Psi^{+}\gamma_{44}$. It corresponds to the duality rotations:
\begin{eqnarray}
\cases{
F_{\mu\nu}\rightarrow F_{\mu\nu}cos\theta+^{\,\,*}F_{\mu\nu}sin\theta,& $ $\cr
^{*}F_{\mu\nu}\rightarrow - F_{\mu\nu}sin\theta+^{\,\,*}F_{\mu\nu}cos\theta,& $ $}
\end{eqnarray}
implemented by Sudbery~\cite{Sudbery}.

Considering the Weinberg ``spinor" in accordance  with Eq. (\ref{eq:EH}) and
restricting
oneself by the first term of Lagrangian (\ref{eq:Lagr})
\footnote{It is possible because in terms of $\vec E$ and $\vec H$ the both of
Eqs. (\ref{eq:LagEul}), obtained from the first and second terms of
(\ref{eq:Lagr}), lead to the same motion equations.}, we get the following
conserved quantities:
\begin{eqnarray}
T_{\{i}^{\quad 4\}4}&=&(\vec E\vec\nabla)\vec H-(\vec H\vec\nabla)\vec E+
\vec E(\nabla\vec H)-\vec H(\vec\nabla\vec E),\\
T_{\{4}^{\quad 4\}4}&=&\vec E [\vec\nabla\times\vec H]-\vec
H[\vec\nabla\times\vec E],\\
T_{\{i}^{\quad j\}4}&=&\vec\nabla\vee\left [\vec E\times\vec H\right ],\\
T_{[i}^{\quad 4]4}&=&-i[(\vec E\vec\nabla)\vec E+(\vec H\vec\nabla)\vec H+
\vec E(\vec\nabla\vec E)+\vec H(\vec\nabla\vec H)],\\
\tilde T_{i}\qquad&=&{1\over 2}\epsilon_{ijk}T_{[j}^{\quad k]4}=\left [(\vec
E\vec\nabla)\vec H-
(\vec H\vec\nabla)\vec E+\vec H(\vec\nabla\vec E)-\vec E(\vec\nabla\vec
H)\right ].
\end{eqnarray}

The value of $A^{\Psi}_{\mu\beta}$ is shown in~\cite{Sankaranarayanan}
to be $A^{\Psi}_{\mu\beta}=-{1 \over 6}\gamma_{5,\mu\beta}$ and,
correspondingly, $A^{\bar\Psi}_{\mu\beta}={1 \over 6}\gamma_{5,\mu\beta}$, (the $S=1$ case).
As opposed to ~\cite{Sudbery} we obtained
\begin{equation}
S^{4,ij}_{4}=0,
\end{equation}
but
\begin{equation}
S^{4,4i}_{4}=-4\left [\vec E\times\vec H\right ]_i.
\end{equation}
At last, we have the same expressions for $J^{\mu}_{\alpha}$ as in
Ref.~\cite{Sudbery}:
\begin{eqnarray}
J_{44}&=&-2(\vec E^2+\vec H^2),\\
J_{4i}&=&4i\epsilon_{ijk}E_j H_k,\\
J_{ij}&=&2[(\vec E^2+\vec H^2)\delta_{ij}-E_i E_j - H_i H_j],
\end{eqnarray}
which are the components of energy-momentum tensor in the common-used
formulation of QED.
Thus, the gauge transformations of the first kind lead to the energy-momentum
conservation and the ``charge" is identified with the energy density of the
field.

The scalar Lagrangian of the Joos-Weinberg's $2(2S+1)$-- theory was presented
in~\cite{Santos,Dvoeglaz} :
\begin{equation}\label{eq:Lagra}
{\cal
L}^{JW}=\partial_{\mu}\bar\Psi\gamma_{\mu\nu}\partial_\nu\Psi+m^2\bar\Psi\Psi.
\end{equation}

Let us note,  implying the interpretation of the Weinberg's 6-``spinor"  as in
(\ref{eq:EH}),
we can rewrite the Lagrangian (\ref{eq:Lagra}) in the following form:
\begin{equation}\label{eq:Lagran}
{\cal L}^{JW}=(\partial_\mu F_{\nu\alpha})(\partial_\mu F_{\nu\alpha}) -
2(\partial_\mu F_{\mu\alpha})(\partial_\nu F_{\nu\alpha}) + 2(\partial_\mu
F_{\nu\alpha})(\partial_\nu F_{\alpha\mu})\,.
\end{equation}
It leads to the Euler-Lagrange equation:
\begin{equation}
{\,\lower0.9pt\vbox{\hrule \hbox{\vrule height 0.2 cm \hskip 0.2 cm \vrule
height 0.2
cm}\hrule}\,}F_{\alpha\beta}-2(\partial_{\beta}F_{\alpha\mu,\mu}-\partial_{\alpha}F_{\beta\mu,\mu})=0,
\end{equation}
where ${\,\lower0.9pt\vbox{\hrule \hbox{\vrule height 0.2 cm \hskip 0.2 cm
\vrule height 0.2 cm}\hrule}\,}=\partial_{\nu}\partial_{\nu}$.
The Lagrangian (\ref{eq:Lagran}) is found out here to be
equivalent to the Lagrangian of the free massless skew-symmetric field
given in~\cite{Ogievetsky,Hayashi}\,\footnote{See also description of closed strings
on the base of this Lagrangian in~\cite{KalbRamond,Love}.}:
\begin{equation}\label{eq:LagHa}
{\cal L}^{H}=\frac{1}{8}F_k F_k,
\end{equation}
with $F_k=i\epsilon_{kjmn}F_{jm,n}$. It can be rewritten
\begin{eqnarray}
{\cal L}^{H}&=&{1\over 4}(\partial_{\mu} F_{\nu\alpha})(\partial_{\mu}
F_{\nu\alpha})-
{1\over 2} (\partial_{\mu} F_{\nu\alpha})(\partial_{\nu}
F_{\alpha\mu})=\nonumber\\
&=&-{1 \over 4}{\cal L}^{JW}-{1\over 2}(\partial_{\mu}
F_{\alpha\mu})(\partial_{\nu} F_{\alpha\nu}),
\end{eqnarray}
which confirms the above statement, taking into account the possibility of 
the Fermi method {\it mutatis mutandis} as in Ref.~\cite{Hayashi}. The second
term in (\ref{eq:Lagran}) can be excluded by means of the generalized Lorentz
condition (which is formally similar to the  well-known Maxwell equations within normalizations of 
the field functions)\footnote{Let us
mention some analogy with the potential formulation of QED. In some sense the
Lagrangian (\ref{eq:Lagran}) corresponds to the choice of
``gauge-fixing" parameter $\xi=-1$, ${\cal L}^{H}$  of  Ref.~\cite[formula
(5)]{Hayashi} corresponds to the ``Landau gauge" ($\xi =0$), and ${\cal L}^{H}$
(formula (9) of cited paper) is in  the ``Feynman gauge"($\xi=1$) for the antisymmetric tensor fields.}.

In  turn the Lagrangian (\ref{eq:LagHa}) is invariant under new ``gauge"
transformations:
\begin{equation}\label{eq:gauge}
F_{\mu\nu}\rightarrow
F_{\mu\nu}+A_{[\mu\nu]}=F_{\mu\nu}+\partial_{\nu}\Lambda_{\mu}-
\partial_{\mu}\Lambda_{\nu}
\end{equation}
The cited paper~\cite{Hayashi} proves that the Lagrangian describes massless
particles
having the longitudinal physical components only. The transversal components
are removed
by means of the ``gauge" transformation (\ref{eq:gauge}).
If we implement this ``gauge" transformations to the ``bivector"
\footnote{See Ref.~\cite[p.244]{Jancewicz} for discussion of Clifford algebra
in the Minkowski space.}:
\begin{equation}
F\rightarrow F+ e_4\wedge A_{[4k]}e_k+{i \over 2} A_{[jk]} e_j\wedge
e_k=F+{1\over 2} A_{[\mu\nu]}e_{\mu}\wedge e_{\nu}
\end{equation}
we can obtain the same result. It is  surprising in the point of view of
the Weinberg theorem about connection between the helicity $\lambda$ and the
Lorentz group representation $(A,B)$,\, namely, $B-A=\lambda$.

Now we turn to the interaction of the $S=1$ partcile in the Joos-Weinberg formalism.
In Ref. [2a, p.B1323] and Ref.~\cite[p.361]{Marinov} the following
invariant (the interaction Hamiltonian)  for interaction of 3-``bispinors" (e.g..
two particles of the spin $S=1/2$ and one particle of the spin $S=1$) has been
constructed:
\begin{equation}\label{eq:hm}
{\cal H}_{\Psi\psi\psi}=g\sum_{\mu_1\,\mu_2\,\mu_3}\left (\matrix{
S_1 & S_2 & S_3 \cr
\mu_1 & \mu_2 & \mu_3 \cr
}\right ) \Phi^{\mu_1}_{(S_1)}\phi^{\mu_2}_{(S_2)}\phi^{\mu_3}_{(S_3)}
\pm \left (\matrix{
S_1 & S_2 & S_3 \cr
\dot{\mu}_1 &\dot{\mu}_2 & \dot{\mu}_3 \cr
}\right )
\Xi^{\dot{\mu}_1}_{(S_1)}\chi^{\dot{\mu}_2}_{(S_2)}\chi^{\dot{\mu}_3}_{(S_3)},
\end{equation}
\\
where
$\left (\matrix{
S_1 & S_2 & S_3 \cr
\mu_1 & \mu_2 & \mu_3 \cr
}\right )$
are the Wigner $3j$- symbols.\\

Assuming the interpretation of the Weinberg's spinor as the sum of vector and
pseudovector\footnote{ In Ref. ~\cite{Cabbibo,Salam}   the
importance of the pseudovector potential $C_k$ in QED has been discussed. In the Singleton papers~\cite{Singleton} as well.}$^{,}\,
$\footnote{As shown in my previous papers the interpretation
$\Psi^{(S=1)}$ according to [2b, p.B888] leads to the contradiction with the
theorem about connection between the $(A,B)$ representation of the Lorentz
group and the helicity of a particle with the field function which transforms according to
this representation ($B-A=\lambda$). Moreover, the Weinberg's massless equations
[2b, formulas (4.21) and (4.22)] admit the acausal ($E\neq\pm p$)
solutions.}:
\begin{eqnarray}
\cases{\chi_{k} = {C}_{k} + iA_{k},& $ $\cr
\phi_{k} = {C}_{k} - iA_{k}\,.& $ $}
\end{eqnarray}
In the case of the massless  helicity-1 particles (photons) we get the following
invariant for interaction of two spinor particles with the generalized electromagnetic
field (the spinor representation is used) :
\begin{eqnarray}\label{eq:lagr}
\lefteqn{{\cal H}_{\Psi\psi\psi}=g\sum_{k\,\mu_2\,\mu_3}\left\{\left [\left
(\matrix{
1 & {1\over 2} & {1\over 2} \cr
k & \mu_2 & \mu_3 \cr
}\right ) \phi^{\mu_2}_{({1\over 2})}\phi^{\mu_3}_{({1\over 2})}
+ \left (\matrix{
1 & {1\over 2} & {1\over 2} \cr
k &\dot{\mu}_2 & \dot{\mu}_3 \cr
}\right )
\chi^{\dot{\mu}_2}_{({1\over 2})}\chi^{\dot{\mu}_3}_{({1\over 2})}\right
] C_k+\right.}\nonumber\\
&+&i\left.\left [ \left (\matrix{
1 & {1\over 2} & {1\over 2} \cr
k & \mu_2 & \mu_3 \cr
}\right )\phi^{\mu_2}_{({1\over 2})}\phi^{\mu_3}_{({1\over 2})}
- \left (\matrix{
1 & {1\over 2} & {1\over 2} \cr
k &\dot{\mu}_2 & \dot{\mu}_3 \cr
}\right )
\chi^{\dot{\mu}_2}_{({1\over 2})}\chi^{\dot{\mu}_3}_{({1\over 2})}\right
]A_k\right\}.
\end{eqnarray}
In (\ref{eq:hm}) we choose the sign $"+"$. The question of the Lorentz transformation rules 
of the pseudovector is related to the transformation rules of 3-rank antisymmetric tensor.
Taken into account the relation between the Pauli $\sigma$-- matrices and the
Clebsh-Gordon coefficients (formula on the p. 65 in~\cite{Akhiezer})
\begin{equation}
\sigma^{\mu}_{\alpha\beta}=-\sqrt{3} C^{{1\over 2}\alpha}_{1\mu{1\over 2}\beta}
\end{equation}
one can rewrite the previous expression (\ref{eq:lagr}) as follows:
\begin{equation}\label{eq:Ham}
{\cal H}_{\Psi\bar\psi\psi}= \frac{g}{\sqrt{6}}\left \{-\bar
\psi\alpha_k\gamma_5 \psi C_k+i\bar\psi\alpha_k \psi A_k\right \}.
\end{equation}
In fact, the coupling constant $g$ is equal to $ie\sqrt{6}$, $e$ is electric
charge in QED,\, $k=1,2,3$. The matrix  $\gamma_5$ has been chosen in the diagonal
form:
\begin{equation}
\gamma_5=\pmatrix{
-1 & 0 \cr
0 & 1 \cr
},
\end{equation}
\begin{equation}
\beta=\alpha_4=\pmatrix{
0 & 1 \cr
1 & 0 \cr
},
\end{equation}
and
\begin{equation}
\vec\alpha=\pmatrix{
\vec\sigma & 0 \cr
0 & -\vec\sigma \cr
}.
\end{equation}
One can see that this interaction Hamiltonian leads to the following equations
from the Hamiltonian (\ref{eq:Ham}):
\begin{equation}\label{eq:Dir}
i\hbar\frac{\partial\psi}{\partial t}=c\vec\alpha\cdot(\vec p-e\vec
A-ie\gamma_5\vec{C})\psi+mc^2\beta\psi,
\end{equation}
which is equivalent to the following system ($c=\hbar=1$) for 2-spinors:
\begin{eqnarray}
\cases{\left [(\vec\sigma\vec p)-e(\vec \sigma\vec
A)+ie(\vec\sigma\vec{C})\right ]\xi+m\eta=E\xi,& $ $\cr
\left [-(\vec\sigma\vec p)+e(\vec\sigma\vec
A)+ie(\vec\sigma\vec{C})\right ]\eta+m\xi=E\eta.& $ $}
\end{eqnarray}
Therefore,
\begin{eqnarray}
\lefteqn{(E^2-m^2)\xi=\left \{\vec p\,^2-e\left [(\vec \sigma\vec p)(\vec
\sigma\vec A)+(\vec \sigma\vec A)(\vec\sigma\vec p)\right ]+\right.}\nonumber\\
&+&\left. ie\left [(\vec\sigma\vec
p)(\vec\sigma\vec{C})-(\vec\sigma\vec{C})(\vec\sigma\vec
p)\right ]+e^2\vec
A\,^2+e^2\vec{C}\,^2+2ieE(\vec\sigma\vec{C})\right \}\xi\,,
\end{eqnarray}
and
\begin{eqnarray}
\lefteqn{(E^2-m^2)\eta=\left \{\vec p\,^2-e\left [(\vec \sigma\vec p)(\vec
\sigma\vec A)+(\vec \sigma\vec A)(\vec\sigma\vec p)\right ]-\right.}\nonumber\\
&-&\left. ie\left [(\vec\sigma\vec
p)(\vec\sigma\vec{C})-(\vec\sigma\vec{C})(\vec\sigma\vec
p)\right ]+e^2\vec
A\,^2+e^2\vec{C}\,^2+2ieE(\vec\sigma\vec{C})\right \}\eta\,.
\end{eqnarray}

We would like to mention that $A_k$, the vector potential, is the compensating
field for the gauge transformation of the second kind, and $C_k$ , the
pseudovector potential, is the compensating field for the chirality gauge
transformation\footnote{See, e.g., Ref.~\cite{Strazhev} for discussion of the
chirality ($\gamma_5$) symmetry of massless fields and neutrino theory of
photons.  As for the generalized gauge transformations, one can find them
in~\cite{Barut, Crawford}.}.
Since  we may assign $E_k=rot \,C_k$ we can see that $\vec{E}=\vec{0}$, and
$\vec{H}=\vec{0}$ in the particular  case~\cite{DVOHJ2}. However, the spectrum is influenced by the term
$\vec{C}$. 

We can implement the new $4\otimes 4$- matrix field corresponding to the
electromagnetic field:
\begin{equation}
\Phi_k=\pmatrix{
A_k-i C_k & 0 \cr
0 & A_k+i C_k  \cr
}
\end{equation}
which is described by the Lagrangian:
\begin{equation}
{\cal L}=\bar\Psi^{(S=1)}\gamma_{\mu\nu}p_{\mu}p_{\nu}\Psi^{(S=1)}=
i\bar\Phi_j\left \{-i\epsilon_{ijk}p_4 p_i\otimes \gamma_5+(\vec
p\,^2\delta_{jk}-p_j p_k)\otimes I\right \}\Phi_k.
\end{equation}
The corresponding dynamical invariants are found from the 
energy-momentum tensor, which is written as following:
\begin{eqnarray}
T_{44}&=&i\bar\Phi_j(\vec p\,^2\delta_{jk}-p_j p_k) \Phi_k,\nonumber\\
T_{\,l4}&=&i\epsilon_{ijk}\bar\Phi_j p_i p_l\otimes\gamma_5\,\Phi_k,\nonumber\\
T_{\,4l}&=&i\epsilon_{ljk}\bar\Phi_j p_4 p_4\otimes \gamma_5
\,\Phi_k-2i\bar\Phi_k p_l p_4\Phi_k+i\bar\Phi_k p_k p_4 \Phi_l+i\bar\Phi_l p_4
p_k\Phi_k,\nonumber\\
T_{\,lm}&=&{\cal L}\delta_{lm}+i\epsilon_{mjk}\bar\Phi_j p_l
p_4\otimes\gamma_5\,\Phi_k - 2i\bar \Phi_k p_l p_m\Phi_k+i\bar\Phi_m p_l
p_k\Phi_k+i\bar\Phi_k p_k p_l\Phi_m.
\end{eqnarray}

The author expresses his gratitude to many colleagues for interest in the works and most helpful discussions.
Particularly, I am grateful to A. M. Cetto (IF-UNAM, M\'exico), S. Baskal (METU, Ankara) and R. Erdem (IIT, Izmir) for insightful discussions  and help during my visits to these institutions.



\begin{thebibliography}{99}
\footnotesize{

\bibitem{Joos} Joos H., 1962 Forts. Phys. {\bf 10} 65.\\[-7mm]

\bibitem{Weinberg} Weinberg S., 1964 Phys. Rev. {\bf 133} B1318; ibid
{\bf 134} B882; ibid {\bf 181} 1893.\\[-7mm]

\bibitem{Weaver} Weaver D. L., Hammer C. L. and Good R. H. Jr, 1964 Phys. Rev. {\bf
135} B241.\\[-7mm]

\bibitem{Wigner} Wigner E. P., 1939 Ann. Math. {\bf 40} 149;
1962 In {\it Group Theoretical Concepts and Methods in Elementary Particle
Physics. Lectures of the Istanbul Summer School of Theoretical Physics. Ed.
G\"ursey~F. Gordon and Breach, 1964}.\\[-7mm]

\bibitem{Sankaranarayanan} Sankaranarayanan A. and Good R. H.,  1965 Nuovo Cim. {\bf
XXXVI} 1303; 1965 Phys. Rev. {\bf 140} B509\\ Sankaranarayanan A., 1965 Nuovo
Cim. {\bf XXXVIII} 889.\\[-7mm]

\bibitem{Good} Shay D. and Good R. H. Jr, 1969 Phys. Rev.{\bf 179} 1410; Krase L.
D., Pao Lu and Good R. H. Jr, 1971 Phys. Rev. D{\bf 3} 1275; Good R. H., 1989 Ann.
Phys.(USA) {\bf 196} 1.\\[-7mm]

\bibitem{Tucker} Tucker R. H. and Hammer C. L., 1971 Phys. Rev. D {\bf 3} 2448.\\[-7mm]

\bibitem{Shay} Shay D., Song H. S. and Good R. H. Jr, 1965 Nuovo Cim. Suppl. {\bf 3}
455.\\[-7mm]

\bibitem{Mathews} Mathews P. M., 1966 Phys. Rev. {\bf 143} 978.\\[-7mm]

\bibitem{Williams} Williams S. A., Draayer J. P. and Weber T. A., 1966 Phys.Rev. {\bf
152} 1207.\\[-7mm]

\bibitem{Santos} Santos F. D., 1986 Phys. Lett. B {\bf 175} 110; Santos F. D. and
van Dam H., 1986 Phys. Rev. C {\bf 34} 250; Amorim A.  and  Santos F. D., 1991
Preprint IFM-9-91 Lisboa;
1992 Phys. Lett. {\bf B297} 31.\\[-7mm]

\bibitem{Ahluwalia} Ahluwalia D. V. and Ernst D. J., 1992 Phys. Lett. B {\bf 287} 18; 1992
Mod. Phys. Lett. A{\bf 7} 1967; 1992 Phys. Rev. C {\bf 45} 3010.\\[-7mm]

\bibitem{Dvoeglaz} Dvoeglazov V. V. and Skachkov N. B., 1984 JINR Communications
R2-84-199 Dubna: JINR; 1987 JINR Communications R2-87-882 Dubna:
JINR; 1988 Sov. J. Nucl. Phys. {\bf 48} 1065.\\[-7mm]

\bibitem{BM} Barut A. O.  , Muzinich I. and Williams D. N., 1963 Phys. Rev. {\bf
130} 442.\\[-7mm]

\bibitem{Ohnuki} Ohnuki Y., 1988 {\it Unitary Representations of the Poincar\'e
Group and Relativistic Wave Equations}. World Sci.  Singapore.\\[-7mm]

\bibitem{Recami} Defaria-Rosa M. A., Recami E. and Rodrigues W. A., 1986 Phys. Lett.
B {\bf 173} 233; ibid {\bf 188} 511(E).\\[-7mm]

\bibitem{Recami2} Majorana E 1928-32 {\it Scientific Manuscripts}, as reported
Recami E., Mignani R. and Baldo M.,  1974 Lett. Nuovo Cim. {\bf 11} 568.\\[-7mm]

\bibitem{Chow} Chow T. L., 1981 J. Phys. A {\bf 14} 2173.\\[-7mm]

\bibitem{Gianetto} Gianetto E., 1985 Lett. Nuovo Cim. {\bf 44} 140 145.\\[-7mm]

\bibitem{Collins} Doughty N. A. and Collins G. P., 1986 J. Phys. A {\bf 19} L887;
1986 J. Math. Phys. {\bf 27} 1639; 1987 J. Math. Phys. {\bf 28} 448.\\[-7mm]

\bibitem{Anderson} Anderson N. and Arthurs A. M., 1978 Int. J. Electron. {\bf 45}
333.\\[-7mm]

\bibitem{Rosen} Rosen J., 1980 Am. J. Phys. {\bf 48} 1071,\\[-7mm]

\bibitem{Sudbery} Sudbery A., 1986 J. Phys. A {\bf 19} L33.\\[-7mm]

\bibitem{Lipkin} Lipkin D. M., 1964 J. Math. Phys. {\bf 5} 696.\\[-7mm]

\bibitem{Fushchich} Fushchich V. I., Krivsky I. Yu. and Simulik V. M., 1987 Preprint
IMAN Ukrainian SSR 87.54 Kiev: IMAN (in Russian).\\[-7mm]

\bibitem{Jancewicz} Jancewicz B., 1988 {\it Multivectors and Clifford Algebra in
Electrodynamics}. World Sci. Singapore.\\[-7mm]

\bibitem{Ogievetsky} Ogievetsky V. I. and Polubarinov I. V., 1968 Sov. J. Nucl. Phys. {\bf 4} 210.\\[-7mm]

\bibitem{Hayashi} Hayashi K., 1973 Phys. Lett. {\bf 44B} 497.\\[-7mm]

\bibitem{KalbRamond} Kalb .M and Ramond P., 1974 Phys. Rev. D {\bf 9} 2273.\\[-7mm]

\bibitem{Love} Clark T. E., Lee C. H. and Love S. T., 1988 Nucl. Phys. {\bf B308} 379.\\[-7mm]




\bibitem{Marinov}Marinov M. S., 1968 Ann. Phys. {\bf 49} 357.\\[-7mm]

\bibitem{Cabbibo} Cabibbo N. and Ferrari E., 1962 Nuovo Cim. {\bf 23}
1147; Candlin D. J., 1965 Nuovo Cim. {\bf 37} 1390; Han M. Y. and
Biedenharn L. C., 1971 Nuovo Cim. A {\bf 2} 544; Mignani R., 1976 Phys. Rev.
D {\bf 13} 2437.\\[-7mm]

\bibitem{Salam}  Salam A., 1966 Phys. Lett. {\bf 22} 683.\\[-7mm]

\bibitem{Singleton} Singleton D., 1995 Int. J. Theor. Phys. {\bf 34} 37; 1996 ibid. {\bf 35} 2419;  
1996 Am. J. Phys. {\bf 64} 452.\\[-7mm]

\bibitem {Akhiezer} Akhiezer A. I. and Berestetskii V. B., {\it Quantum
Electrodynamics}. Interscience Publisher, translated by  Volkoff G. M. ,
1965.\\[-7mm]


\bibitem{Strazhev} Strazhev V. I., 1977 Int. J. of Theor. Phys. {\bf 16}
111;  Strazhev V. I. and Kruglov S. I., 1977 Acta Phys. Polon. {\bf B8}
807.\\[-7mm]

\bibitem{Barut} Barut A. and McEwan J., 1984 Phys. Lett. B {\bf 135}
172.\\[-7mm]

\bibitem{Crawford} Crawford J. P., 1993 {\it The Dirac Oscillator and Local
Automorphism Invariance}. Preprint.\\[-7mm]

\bibitem{DVOHJ2} Dvoeglazov V. V., 1993 Hadronic J. {\bf 16} 423.\\[-7mm]

}

\end{thebibliography}
\end{document}